\documentclass[aps, prd, showpacs, superscriptaddress, 
nofootinbib, twocolumn,floatfix,preprintnumbers
]{revtex4} 
\usepackage{graphicx}\usepackage{amsmath}\usepackage{natbib}
\usepackage{times}

\begin{document}

\title{Neutrino Spectrum from SN~1987A and from Cosmic Supernovae}

\author{{Hasan Y{\"u}ksel}}
\affiliation{Department of Physics, Ohio State University, Columbus, Ohio 43210}
\affiliation{Center for Cosmology and Astro-Particle Physics, Ohio State University, Columbus, Ohio 43210}

\author{{John F. Beacom}}
\affiliation{Department of Physics, Ohio State University, Columbus, Ohio 43210}
\affiliation{Department of Astronomy, Ohio State University, Columbus, Ohio 43210}
\affiliation{Center for Cosmology and Astro-Particle Physics, Ohio State University, Columbus, Ohio 43210}

\date{18 October 2007}

\begin{abstract}
The detection of neutrinos from SN~1987A by the Kamiokande-II and Irvine-Michigan-Brookhaven detectors provided the first glimpse of core collapse in a supernova, complementing the optical observations and confirming our basic understanding of the mechanism behind the explosion.  One long-standing puzzle is that, when fitted with thermal spectra, the two independent detections do not seem to agree with either each other or typical theoretical expectations. We assess the compatibility of the two data sets in a model-independent way and show that they can be reconciled if one avoids any bias on the neutrino spectrum stemming from theoretical conjecture. We reconstruct the  neutrino spectrum from SN~1987A directly from the data through nonparametric inferential statistical methods and present predictions for the Diffuse Supernova Neutrino Background based on SN~1987A data. We show that this prediction cannot be too small (especially in the 10-18 MeV range), since the majority of the detected events from SN~1987A were above 18 MeV (including 6 events above 35~MeV), suggesting an imminent detection in operational and planned detectors. 
\end{abstract}

\maketitle

%
\section{INTRODUCTION}\vspace*{-0.4cm}
Core-collapse supernova explosions are how massive stars end their lives~\cite{Heger:2002by}.  Once their nuclear fuel is exhausted, they produce a neutron star or black hole, recycle their ashes into cosmic dust to seed subsequent starbursts and spread their heavy elements~\cite{Yoshida:2006sk,Pruet:2004vb,Pruet:2005qd}, and flood the universe with a burst of neutrinos.  Supernova explosions are very prolific neutrino producers, since the enormous amount of gravitational binding energy, liberated when the core of a massive star collapses, can effectively be disseminated only by neutrinos.  In simulations, the rebounding collapse of the core often produces a shock that stalls, which may indicate the necessity of shock revival by energy transfer from the intense neutrino flux~\cite{Keil:2002in,Burrows:2005dv,Scheck:2006rw,Thompson:2002mw,Kotake:2005zn,Sumiyoshi:2005ri,Fryer:2003jj,Liebendoerfer:2003es,Totani:1997vj}.  While neutrinos can be crucial for diagnosing a successful explosion and revealing the conditions of the core, their elusive nature  makes their detection challenging. 

The detection of neutrinos from SN~1987A in the Large Magellanic Cloud (at a distance of $D\sim50$~kpc) by the Kamiokande-II (Kam-II)~\cite{Hirata:1987hu,Hirata:1988ad} and Irvine-Michigan-Brookhaven  (IMB)~\cite{Bionta:1987qt,Bratton:1988ww} detectors transformed supernova  physics from purely hypothesis- and simulation-driven to discovery-driven science.  One puzzling feature of the SN~1987A data is that the neutrinos detected by the IMB detector were seemingly more energetic than those detected by the Kam-II detector, which were clustered at low energies. Despite the uncertainties associated with the low statistics and the differences of the detector properties (like efficiency and size), it seems hard to accommodate the two detections with a common quasi-thermal neutrino spectrum~\cite{Jegerlehner:1996kx}. This incompatibility between the data sets and the theory can be alleviated by assuming that at least one experiment had an anomalous statistical fluctuation or underestimated systematics~\cite{Costantini:2006xd}. Here, we assume that both experiments observed statistically probable outcomes, as they were exposed to the same spectrum, and accept the reported events, uncertainties and overall efficiencies at face value.

The principal focus of our study is to assess the compatibility of the Kam-II and IMB data sets in a model-independent fashion, and to study the immediate implications for neutrino detection from Galactic and cosmic supernovae.  We begin with a brief formulation of neutrino detection from a nearby supernova and summarize the main ingredients.  Next, we determine the relative prospects of detecting neutrinos from SN~1987A in either Kam-II or IMB, based on their reported energy-dependent efficiencies and the numbers of targets in their fiducial volumes. We compare this relative probability of detection to the actual detected positrons for Kam-II and IMB and conclude that there is no significant discrepancy between the data sets unless an a priori spectral shape is forced to fit  the two data sets simultaneously. We establish that the Kam-II and IMB detectors mainly probed different energy domains, the former testing the low energy part of the spectrum and the latter testing the high energy part, and, in fact, only a proper combination of Kam-II or IMB data sets provides a sensitive probe over all energy ranges. 

Using nonparametric (distribution-free) inferential statistical methods~\cite{Silverman:1900}, we determine the structure of the underlying spectrum directly from the data. Since these methods make no a priori assumptions and do not rely on parameter estimation, they allow for more efficient processing of small data samples. Then, we present the incoming neutrino spectrum inferred directly from the data, which would be an effective spectrum received on Earth after effects modifying the overall shape are taken into account. Finally, we provide predictions for the Diffuse Supernova Neutrino Background based on the SN~1987A data (as Refs.~\cite{Fukugita:2002qw,Lunardini:2005jf} have also recently done through more conventional methods), which turn out to be nearly as large as typical predictions in the literature based on supernova models. Our study follows a transparent approach to extract the targeted information directly from the SN~1987A data, skipping the intermediate stages of multiparameter function fitting, commonly used in the past literature. The methods we present avoid over-interpretation of limited data, while still allowing for a substantial analysis.

\section{Detection of Supernova Neutrinos}  %
We concentrate on the $\bar\nu_e$ flux received on Earth from a supernova since the detectors are particularly sensitive to the inverse beta decay, $\bar\nu_e+p\to n+e^+$ (we ignore possible small contributions due $\nu+e^-$ or $\nu + ^{16}$O). We parametrize the $\bar\nu_e$ number spectrum as $\phi (E_\nu) = [{L_\nu}/{E_0}] \, \varphi (E_\nu)$, where $\varphi (E_\nu)$ is a normalized function ($\int\varphi(E_\nu) \,dE_\nu=1$) describing the overall spectral shape with no prior assumptions.  Here $E_0$ is the average energy ($\langle E_\nu\rangle =  \int E_\nu \varphi(E_\nu) \, dE_\nu = E_0$) and $L_\nu$ is the time-integrated luminosity (i.e., the total number of neutrinos  is ${L_\nu}/{E_0}$) in this flavor.  In the energy regime of  interest, the cross section of inverse beta decay at the lowest orders in inverse nucleon mass, ${1/M}$, is
\begin{equation}
\sigma(E_\nu) \simeq 9.5\times 10^{-44} {\mathrm{cm}^2}  \left(1 - 6 \, {E_\nu}/{M}\right) \left[{(E_\nu -\Delta)}/{\mathrm{MeV}}\right]^2 ,
\end{equation}
expressed in terms of the neutrino energy, $E_\nu$~\cite{Vogel:1999zy}. The detected positron energy (visible energy), $E_+$, can be related to the incoming neutrino energy as: $E_{+} = E_\nu - \Delta$, where $\Delta ={M}_{n} - {M}_{p}$ is the nucleon mass difference (the higher order corrections are much smaller than uncertainties associated with energy resolution).  The positron spectrum in a given detector, as a function of visible energy for a supernova at a distance $D$, can be cast as
\begin{equation}
\Phi(E_{+}) =  \frac{N(E_{+})}{4 \pi D^2} \sigma(E_\nu)\, \phi (E_\nu) ,
\label{Eq:nu_flux}
\end{equation}
where the effective number of targets available in the detector is $N(E_+)=N_t \eta(E_{+}) $ as a function of visible (positron) energy, in which  $N_t$ stands for the number of free-proton targets weighted with the up-time fraction of the detector and $\eta(E_{+}) $ is the detector efficiency function. 

\section{Compatibility of the Data Sets}
\begin{figure}[b]
\includegraphics[width=3.4in,clip=true]{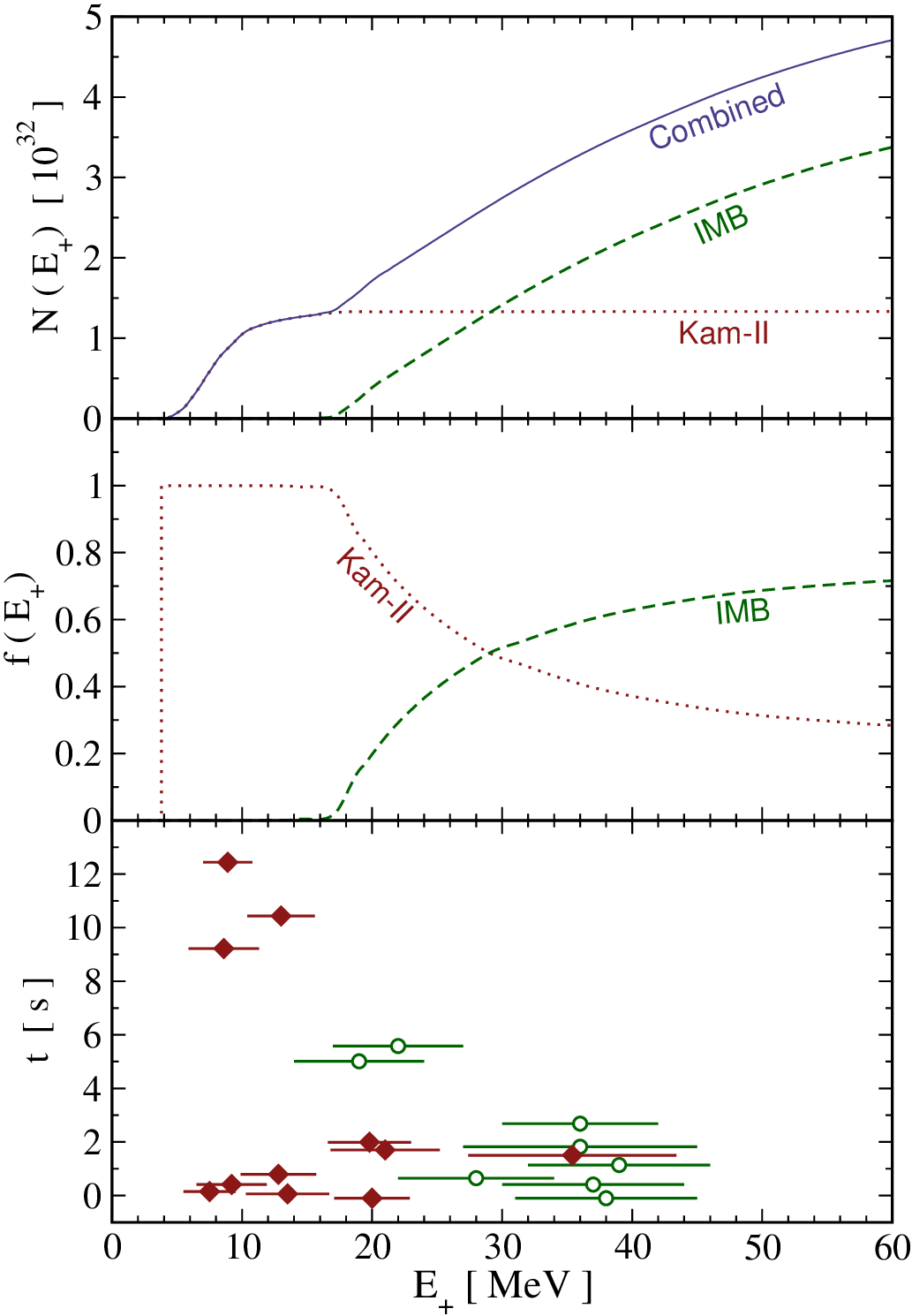}
\caption{\label{Fig:1} 
\textbf{Top Panel:} The effective number of targets (free protons) available in the Kam-II and IMB detectors and their combination as a function of the visible (positron) energy~$E_{+}$ (i.e., protons in the fiducial volume of each detector normalized by the up-time and energy-dependent efficiencies).  
\textbf{Middle Panel:} The fraction of total detected events expected to be assigned to either Kam-II or IMB (independent of incoming neutrino spectrum) as a function of the visible energy.
\textbf{Bottom Panel:} Positrons detected during SN~1987A by Kam-II (11 filled-diamonds) and IMB (8 open-circles) with their recorded energies ($E_i$) and experimental uncertainties ($\Delta E_i$). }
\end{figure}
In the top panel of Fig.~\ref{Fig:1}, we present the effective number of targets available in the Kam-II and IMB detectors by normalizing the targets in the fiducial volume of each detector by the up-time and reported energy-dependent efficiencies~\cite{Burrows:1988ba,Loredo:2001rx}, $ N(E_+) = N_t \, \eta(E_{+})$. The combined effective number of targets for both detectors together is also shown (in which they are treated as one big detector with energy-dependent efficiency, covering  the full energy range sampled by the data). The number of free-proton targets in the fiducial volume of each detector was $ N_t \simeq 1.43 \times 10^{32}$ for Kam-II and $N_t \simeq 4.55 \times 10^{32}$ for IMB~\cite{Burrows:1988ba}.  Despite being $\sim3$ times larger in size, IMB typically had lower efficiency, especially at lower energies. Moreover, the IMB detector was briefly offline whenever a muon passed through the detector, reducing the average up-time by $\sim$13\%~\cite{Jegerlehner:1996kx}. We emphasize that while all previous analyses did take the separate energy-dependent efficiencies of Kam-II and IMB into account, our approach of considering an effective combined detector with a more complicated energy-dependent efficiency is new.

Note that the efficiency of Kam-II (and most modern neutrino detectors) is basically a step function over the detector threshold. Kam-II had a low threshold of $E_{thr} \sim 7.5$~MeV and its efficiency quickly rose, starting at $\sim 5$~MeV, to its asymptotic value. This is because the efficiency function around the threshold energy is non-trivial due to the probabilistic nature of the detection process (i.e., in principle, a low energy neutrino could trigger a greater number of photomultipliers and be detected).  At the time of SN~1987A, almost a quarter of the photomultipliers in the IMB detector were offline due to a power failure, resulting in a distinct energy dependence due to the geometrical modifications. Thus, IMB had a higher threshold of $E_{thr} \sim 19$~MeV and rose gradually, starting at $\sim 16$~MeV. At low energies, the yields will be dominated by Kam-II, due to its much greater efficiency, while at high energies, the yields will be dominated by IMB, due to its greater size.

When both the efficiencies and numbers of target protons in the fiducial volumes of the two detectors are taken into account simultaneously, it becomes apparent why Kam-II registers all the positrons below 16~MeV, while IMB is expected to dominate over 30~MeV. An informative perspective in  comparing the effectiveness of Kam-II and IMB is presented in the middle panel of Fig.~\ref{Fig:1}. This shows the fraction of total positrons expected to be recorded by either of the experiments (i.e., the ratio of effective number of targets in each detector to the combined number of targets) as a function of visible energy. Note that this  is independent of the shape of the incoming neutrino spectrum or the cross section, since it only compares the effectiveness of the detectors relative to each other.  One can directly compare this information to the actual positrons detected during SN~1987A in order to assess the compatibility of the detectors, without ever referring to the actual received spectrum of neutrinos. As explained below, we are assuming a common incoming neutrino spectrum.

We summarize the visible energies ($E_i$) and experimental uncertainties ($\Delta E_i$) of positrons detected at the time of SN~1987A in Kam-II (11 data points marked with filled-diamonds) and IMB (8 data point marked with open-circles) in the bottom panel of Fig.~\ref{Fig:1}. We exclude the Baksan Scintillator Telescope data~\cite{Alekseev:1987ej,Alekseev:1988gp},  which has much larger uncertainties associated with backgrounds, from our main results but comment on possible impact later. Data sets are synchronized based on the arrival time of the first events (note that we limit our study to time-integrated neutrino flux). While Kam-II originally reported 12 signal events, one of them, being attributed to background, is excluded from our analysis, as in previous studies. The visible energy assigned to each detected event is based on the number of triggered photomultipliers from the \v{C}erenkov light of the positrons. The statistical uncertainty on this visible energy is proportional to the square root of the total number of hits and is shown by a horizontal bar. 

Interestingly, the distributions of the detected positrons between Kam-II and IMB, as shown in the bottom panel, quantitatively agrees quite well with our  expectations based on the middle panel of Fig.~\ref{Fig:1}. While  Kam-II dominated the detections below 16~MeV, the events between 16--30~MeV in each detector are comparable  as one would deduce from the middle panel. At even higher energies (where the direct comparison is possible), IMB detected 5 events while Kam-II detected only 1. If we accept that the IMB observation of 5 events is a statistically probable outcome, since the effective number of targets in IMB is almost twice to that of Kam-II at these energies, one would expect to see $\sim 2.5$ events in Kam-II.  With this expectation, the probability of seeing $\lesssim 2$ events is over 50\%, so that the observation of only a single event in the Kam-II sample at high energies is probably simply a downward fluctuation, and not in disagreement with the IMB sample. Instead, if we assume that the Kam-II observation of 1 event is correct, then one would expect to see $\sim 2$ events in IMB. With such a low expectation, seeing 5 events in IMB is quite unlikely (with a 15\% probability of getting 4 or more events, and a 4\% probability of getting 5 or more events), suggesting instead that the true spectrum was indeed most faithfully sampled by the larger IMB detector.

Considering the low statistics and large experimental uncertainties involved, it seems that there is no obvious conflict between the distribution of detected positrons in the two experiments, suggesting the possibility of reconciling the Kam-II and IMB data sets with a suitable incoming neutrino spectrum. It is evident that while each experiment may suggest a radically different incoming neutrino spectrum individually, since the signal will be sampled distinctly by the detectors, only their proper combination could provide a sensitive probe over all energy ranges. In order to minimize the impact of statistics due to the small sample, we will present our subsequent results with an emphasis on the combined data.

\section{Expected Neutrino Spectrum}
Neutrinos are produced thermally from the plasma and remain trapped inside the collapsed core of a massive star until their optical depth becomes unity around the corresponding neutrino-sphere of each flavor. Then they stream  through the envelope, preserving their initial energy distribution. The average energy at free streaming is dictated by the temperature at the neutrino-sphere, which is in turn determined by the strength of the coupling of neutrinos to the matter. One expects a hierarchy of energies, $ \langle E_{\nu_e}  \rangle < \langle E_{\bar \nu_e}\rangle < \langle E_{\nu_{\mu, \bar{\mu}}}\rangle$, since the electron neutrino flavor enjoys both charged and neutral current interactions (and couples to matter most strongly), while the non-electron flavors feel a weaker attachment to the plasma. The time-integrated luminosity and the hierarchy of energies among neutrino flavors from a core-collapse supernova are still not well known and diverse values are reported by modelers~\cite{Keil:2002in,Burrows:2005dv,Scheck:2006rw,Thompson:2002mw,Kotake:2005zn,Sumiyoshi:2005ri,Fryer:2003jj,Liebendoerfer:2003es,Totani:1997vj}. While the  supernova neutrino spectra are expected to be quasi-thermal, modifications due to non-standard effects, like neutrino mixing among various flavors~\cite{Walker:1986xd,Takahashi:2001ep,Minakata:2001cd,Lunardini:2004bj,Fogli:2004ff,Tomas:2004gr,Fuller:1998kb,Beun:2006ka}, neutrino decay~\cite{Beacom:2002cb,Ando:2004qe,Fogli:2004gy,Ando:2003ie,Lindner:2001th}, neutrino-neutrino interactions~\cite{Balantekin:2004ug,Duan:2006jv,Hannestad:2006nj,Balantekin:2006tg}, additional channels of energy exchange between flavors~\cite{Keil:2002in,Raffelt:2001kv,Horowitz:2003yx,Carter:2001kw}, and/or any other novel mechanism due to unknown physics, may produce a time-integrated spectrum received on Earth that deviates significantly from a quasi-thermal shape. 

Many previous studies dealing with the SN~1987A neutrinos adopt a template neutrino spectrum and try to extract parameters describing this spectrum from the Kam-II and IMB data. Both Maxwell-Boltzmann~\cite{Bahcall:1987ua}, $\varphi(E_\nu)\propto {E_\nu^2}{e^{-E_\nu/T}} $,  and  Fermi-Dirac spectra~\cite{Krauss:1987re}, $\varphi(E_\nu) \propto {E_\nu^2}/[e^{E_\nu/T}+1]  $, require only a single temperature parameter, $T$. An additional degeneracy parameter, $\eta$, is used to describe spectral pinching in a Fermi-Dirac distribution~\cite{Janka89}, $\varphi(E_\nu) \propto {E_\nu^2}/[e^{E_\nu/T-\eta}+1]$. More recently, a quasi-thermal distribution (based on a gamma distribution), is suggested~\cite{Keil:2002in}, $ \varphi(E_\nu) \propto E_\nu^\gamma  \, e^{-(\gamma+1) {E_\nu}/{E_0} } $, which can handle anti-pinched (broader) spectra, as well as pinched ones, through the parameter $\gamma$. It has been shown that SN~1987A data may favor a highly anti-pinched spectrum~\cite{Mirizzi:2005tg}. However, all such pre-defined functions describe only a very limited class of possible shapes, which may not necessarily fit the actual received spectrum of neutrinos.

More complex composite spectral shapes are also considered in the literature.  Today, it is well established that neutrinos may change flavor while they travel through  matter, so-called MSW effect~\cite{Wolfenstein:1977ue, Mikheev:1986gs, Balantekin:2003dc,Bahcall:2004ut,Maltoni:2004ei,Balantekin:2003jm,Fogli:2005cq}. The density profile of the supernova envelope, which neutrinos travel through, spans many orders of magnitude and thus enables many resonances and opportunities to swap flavors. The IMB and Kam-II detectors are expected to observe almost identical spectra since the Earth effects~\cite{Akhmedov:2002zj, Takahashi:2001dc,Dighe:2003vm} are unlikely to introduce significant differences (using recent determinations of the mixing parameters). A superposition of  low-energy/high-luminosity and  high-energy/low-luminosity thermal spectra due to neutrino mixing fits the data better than a single quasi-thermal spectrum~\cite{Lunardini:2004bj}.  A bimodal neutrino distribution theoretically arising from the accretion and the cooling phases of a supernova~\cite{Loredo:2001rx}, is another example of a composite spectrum.  However, due to limited statistics, inferring multiple theoretical parameters is a challenging task, due to severe degeneracies among them. While both approaches effectively increase the number of parameters describing the spectrum, naturally improving the fit to the data, one can only probe an effective ${\bar \nu_e}$ spectrum after any physical   mechanisms modifying the spectrum, including oscillations, are taken into to account.

\section{Neutrinos from SN~1987A}
\begin{figure}[b]
\includegraphics[width=3.4in,clip=true]{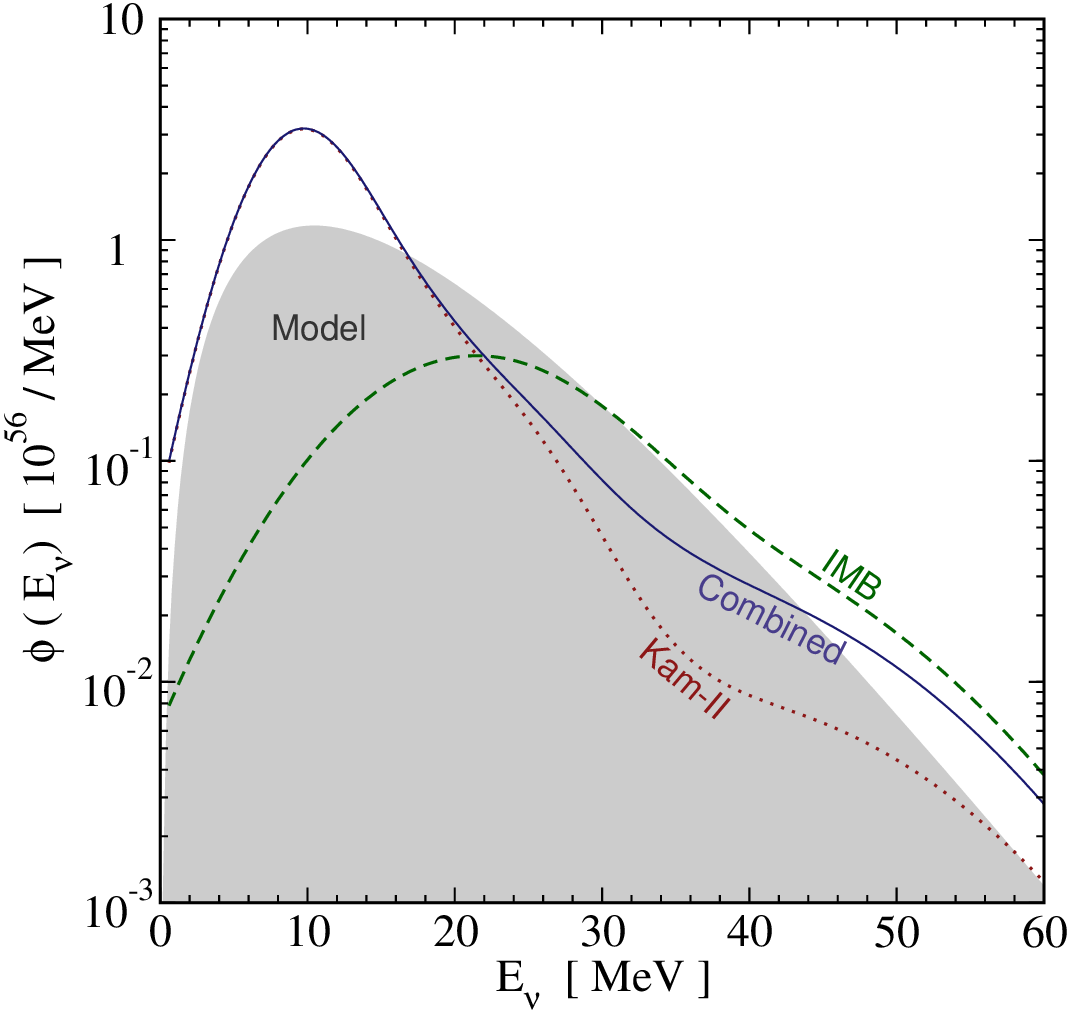}
\caption{\label{Fig:2} The inferred neutrino emission spectra from either the Kam-II or IMB data sets alone or their combination, as discussed in the text (taking into account the corresponding effective number of targets and cross section). The shaded shape is a Fermi-Dirac spectrum with canonical neutrino emission parameters (average energy $E_0 = 15$~MeV and time-integrated luminosity $L= 5\times 10^{52}$~erg).}
\end{figure}
We resort to the data directly and attempt to find the simplest description of the effective ${\bar \nu_e}$ spectrum that could be useful for studies which may require such input, regardless of the underlying physics and assumptions. As noted, we assume that the two detectors were exposed to the same incoming neutrino spectrum. The theoretical relation between the received neutrino spectrum $\phi (E_\nu)$ and the corresponding positron detection spectrum $\Phi(E_{+})$ is given in Eq.~(\ref{Eq:nu_flux}). The actual detected positron spectrum as recorded by an experiment can be expressed as a sum of ``bumps'' placed at the observations, $ \Phi(E_{+}) =  \sum_i \delta(E_{+}-E_i)$, where the index is over the set of events under consideration. This would be a faithful representation of the theoretically-expected positron spectrum (assuming that the data was a probable outcome and not a statistical anomaly), only if the detected number of positrons were very large (i.e., in the case of a future Galactic supernova, operational detectors are expected to observe many thousands of events) and the uncertainties of the visible energies were small. Then, one can infer the received neutrino spectrum by inverting Eq.~(\ref{Eq:nu_flux}) as
\begin{equation}
\phi (E_\nu)=  {4 \pi D^2} \sum_i \frac{\delta(E_+-E_i)}{ N(E_{i}) \, \sigma(E_{i} +\Delta) }\, ,
\label{eq:phi}
\end{equation}
which can be verified by substituting back into Eq.~(\ref{Eq:nu_flux}). When the statistics are limited, as here, we must regulate the discrete bumps in the spectrum by an appropriate smoothing; the most physically motivated method is to use the approximately Gaussian uncertainties on the measured energies. Hence we replace each $\delta$ function by a Gaussian with a  width of $\varepsilon_i$,
\begin{equation}
\delta(E_+-E_i) \rightarrow {\exp[{ -(E_{+}-E_i})^2 /(2 \varepsilon_i^2)]}/(\sqrt{2\pi} \varepsilon_i) \,.
\label{eq:delta}
\end{equation}
Since  very sharply peaked Gaussians would reveal the spurious fine structure of the data, we choose a generous width of $\varepsilon_i= 1.5 \Delta E_i$ (where $\Delta E_i$ is the uncertainty on the  assigned energy of each detected positron), which enables us both to take into account uncertainties associated with the detection process adequately and to present a sufficiently smooth spectrum. Any larger width would introduce excessive smoothing, which would obscure details of the distribution and spuriously enhance the tails. Note that in order to directly reconstruct the neutrino spectrum in this way, it is crucial that the neutrino and positron energies are related in a one-to-one way as in inverse beta decay (unlike for neutrino-electron scattering, for example).

Figure~\ref{Fig:2} displays the inferred spectra from either the Kam-II or IMB data sets, as a function of neutrino energy.  The differences between the  inferred spectra from Kam-II or IMB stems from a few facts. Since the IMB detector has no sensitivity at lower energies, where we can rely on Kam-II only, the two least energetic events of IMB, for which the efficiency is very low, then become disproportionately significant. Also the fact that only a single positron was detected by Kam-II at higher energies requires a significant suppression of the neutrino spectrum. However, the effective number of targets for  Kam-II is low compared to IMB (as shown in the top panel of the Fig.~\ref{Fig:1}) and the expected number of events could easily fluctuate down.
 
We have earlier established that Kam-II and IMB probed different energy domains and also showed that the detected positrons showed no obvious conflict with this expectation. So rather than focusing on the differences between the spectra suggested by the Kam-II or IMB data sets, we will focus on what can be learned from the combined data set, since both experiments were measuring the identical incoming neutrino spectrum using an identical (water-\v{C}erenkov) technique.

By combining the two data sets and the corresponding (energy dependent) effective number of targets (as displayed in the top panel of Fig.~\ref{Fig:1}), we avoid over-emphasizing the differences due to low statistics. The combined result is also less susceptible to Poisson fluctuations associated with such limited statistics, yet covers the whole energy domain, and is more conservative.  The neutrino spectra based on this combined data set is shown in Fig.~\ref{Fig:2} (solid line).  For comparison, we also show a Fermi-Dirac spectrum with canonical neutrino emission parameters (an average energy $E_0 = 15$~MeV and a integrated luminosity $L_\nu = 5\times 10^{52}$~erg). 

Since we construct the spectra directly from the data, it is not meaningful to make goodness-of-fit tests between the data and these constructed spectra (weighted with the energy-dependent proportionality factors).  If we use delta functions to represent the data, then the cumulative distributions of the measured and `predicted' spectra would be identical, so that a Kolmogorov-Smirnov test would indicate perfect agreement.  When we use Gaussians to represent the data, the Kolmogorov-Smirnov test, while not strictly meaningful, provides confirmation that our chosen width is not too large. 

While a pinched template spectrum  puts more weight to the peak, and an anti-pinched spectrum puts more weight to the tail, the shape from the combined  data can be best explained by a spectral shape that is enhanced both at the peak (to accommodate events from Kam-II where IMB was not sensitive)  and high energy tail of the spectrum (to better accommodate energetic events from IMB), and depressed in between, compared to a thermal Fermi-Dirac spectrum. This basic shape of the underlying spectrum (see our Fig.~\ref{Fig:2}), in agreement with the two-component composite spectrum of Lunardini and Smirnov~\cite{Lunardini:2004bj,Lunardini:2005jf}, could reconcile the Kam-II and IMB data with each other. The corresponding luminosity for the combined spectrum, $L_\nu \sim 6\times 10^{52}$~erg, is quite similar to that of the model, with an average energy, $E_0 \sim 12$~MeV, that is somewhat lower. 

Since we consider a data set of 19 detected positrons, the overall uncertainty due to the Poisson nature of the detection will not be too large, $1/\sqrt{19}\sim 25$\%. In the peak region, our reconstruction is based on 12 events with a nominal uncertainty of $1/\sqrt{12}\sim 30$\%, so that the excess relative to the model is significant. In the tail region, there are 7 events with a nominal uncertainty of $1/\sqrt{7}\sim 40$\%, which is also illustrated by the differences between the spectra in this region of joint sensitivity. We emphasize that the tail region is not so uncertain, as there were 6 events above 35~MeV; this strongly precludes any hypothesized suppression of the tail. In Fig.~\ref{Fig:2}, the combined spectrum above 40 MeV depends on the width chosen for the Gaussians, as there were no events at these energies (however, there were 6 events between 35 and 40 MeV, each with relatively large energy uncertainties, and their statistical weights must go somewhere).  This is the least certain part of the spectrum, due to the low flux there.

\section{DSNB Directly from SN~1987A}

Neutrinos from past core-collapse supernovae have been studied extensively~\cite{Fukugita:2002qw,Lunardini:2005jf,Guseinov,Bisnovatyi-Kogan,Krauss:1983zn, Domogatskii,Dar:1984aj,woosley:1986xxxx, Totani:1995rg, Malaney:1996ar, Hartmann:1997qe, Kaplinghat:1999xi, Ando:2002ky,Beacom:2005it,Strigari:2005hu,Yuksel:2005ae,Ando:2004hc,Daigne:2005xi,Iocco:2004wd} and experimental limits are already suggesting an impending detection~\cite{Malek:2002ns, Eguchi:2003gg, Aharmim:2006wq}. The Diffuse Supernova Neutrino Background (DSNB) flux depends not only on the typical supernova neutrino emission spectrum, but also the expansion rate of the universe (redshift-time relation $h(z)=\sqrt{ \Omega_{M} (1+z)^{3} + \Omega_{ \Lambda} }$, where $\Omega_{M} = 0.3$, $\Omega_{\Lambda} = 0.7$) and the core-collapse supernova rate (SNR), which presumably tracks the history of star formation rate (SFR) (see e.g.~\cite{ Hopkins:2006bw} and references therein).  The neutrino emission per supernova is the most uncertain quantity~\cite{Yuksel:2005ae,Hopkins:2006bw} and is hence our focus here. The current precision of the data shows that the evolution of the SFR can be parametrized with a piecewise linear fit:
\begin{eqnarray}
{  R}_{{  SF}} (z) 
& = &{  R}_{  SF}^{0} \, (1+z)^{\alpha } \quad \, \, \, {  for}
\quad  z < z_{  p} \nonumber\\ 
& = &  \frac{{  R}_{  SF}^{0}   (1+z)^{\beta }}{(1+z_{  p})^{\beta-\alpha}}
\quad {  for}  \quad  z_{p} < z < z_{max}  ,
\end{eqnarray}
where $\alpha \sim 3.44$, $\beta\sim0$, $z_{p}\sim1$ and $z_{max}\sim5$. The overall normalization in the local universe is $ {R}_{SF}^0= 0.0095 M_{\odot}/ (\mathrm{yr \,\, Mpc}^3)$. The core-collapse supernova rate as a function of redshift is ${R}_{SN} (z)= \zeta {R}_{SF} (z)$, while the fraction of stellar mass ending as supernovae, $\zeta = 0.0132 / M_{\odot}$ (for the  Baldry--Glazebrook  IMF~\cite{Baldry:2003xi}), depends on the stellar mass function. We note that this dependence on the assumed stellar mass function is small, as explained in Ref.~\cite{ Hopkins:2006bw}. While the directly measured core collapse supernova rate is slightly smaller~\cite{Dahlen:2004km,cappellaro:2005zzz}, recent studies suggest this may be misleading. A large fraction of supernova exploding in very dusty starburst environments may go undetected, causing a 30-60\% underestimate of the true core-collapse supernova rate~\cite{Mannucci:2007ex}.  Thus, while the supernova rate data are generally confirming, we use the more reliable star formation rate data for now.  Ultimately, it will be possible to use the supernova rate data to more directly predict the DSNB flux, as first pointed out by Ref.~\cite{Strigari:2005hu}.  The uncertainty on the flux due to astronomical inputs is small and will be further reduced with anticipated improvements in the data. In our analysis, we assumed that the normalization of the SFR and its evolution can be determined independently by astronomers, and used as a fixed input to extract more accurate information on typical supernova properties. 

\begin{figure}[t]
\includegraphics[width=3.4in,clip=true]{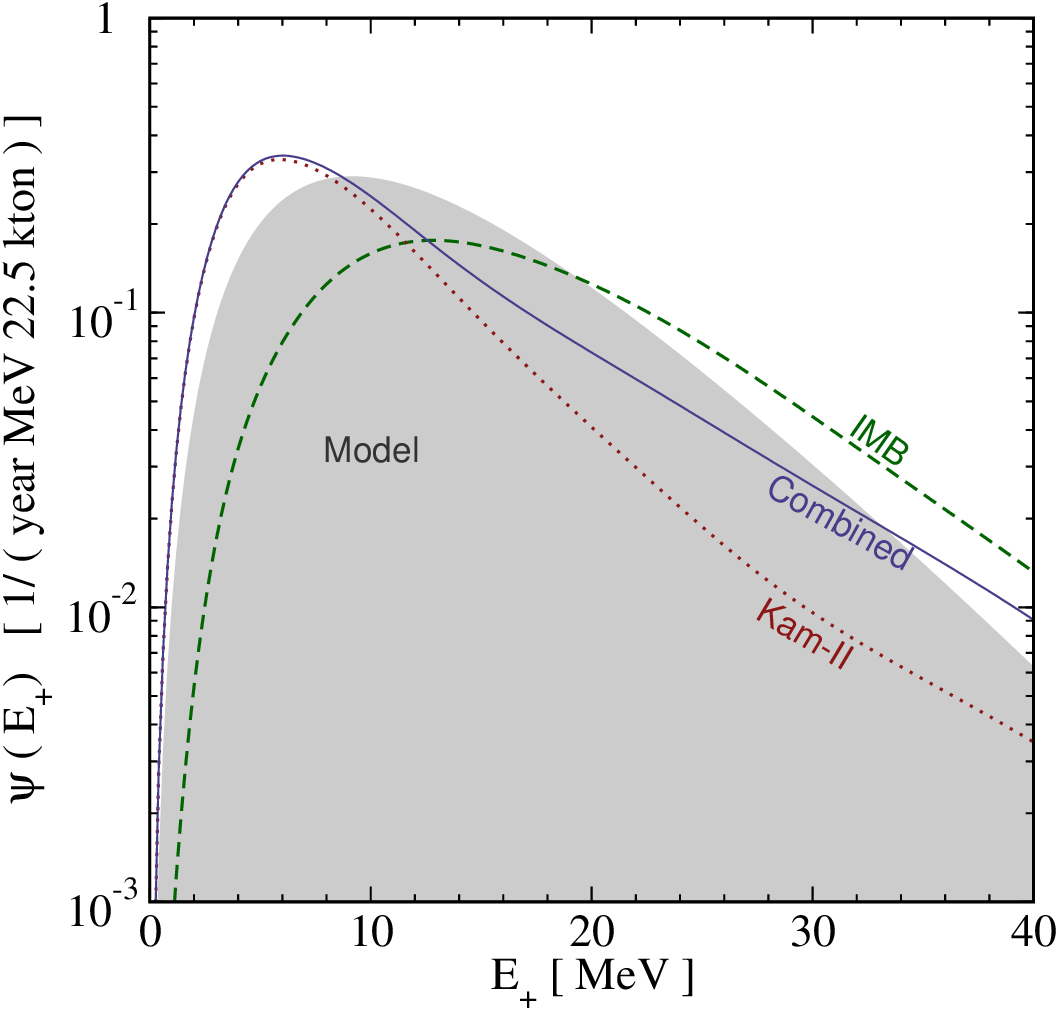}
\caption{\label{Fig:3}
The DSNB detection spectra based on the neutrino spectra inferred from either the Kam-II or IMB data sets alone or their combination  as in Fig.~\ref{Fig:2}, compared to a model (shaded shape) with canonical neutrino emission parameters (the assumed core-collapse SN history is described in the text).
}
\end{figure}
\begin{figure}[b]
\includegraphics[width=3.4in,clip=true]{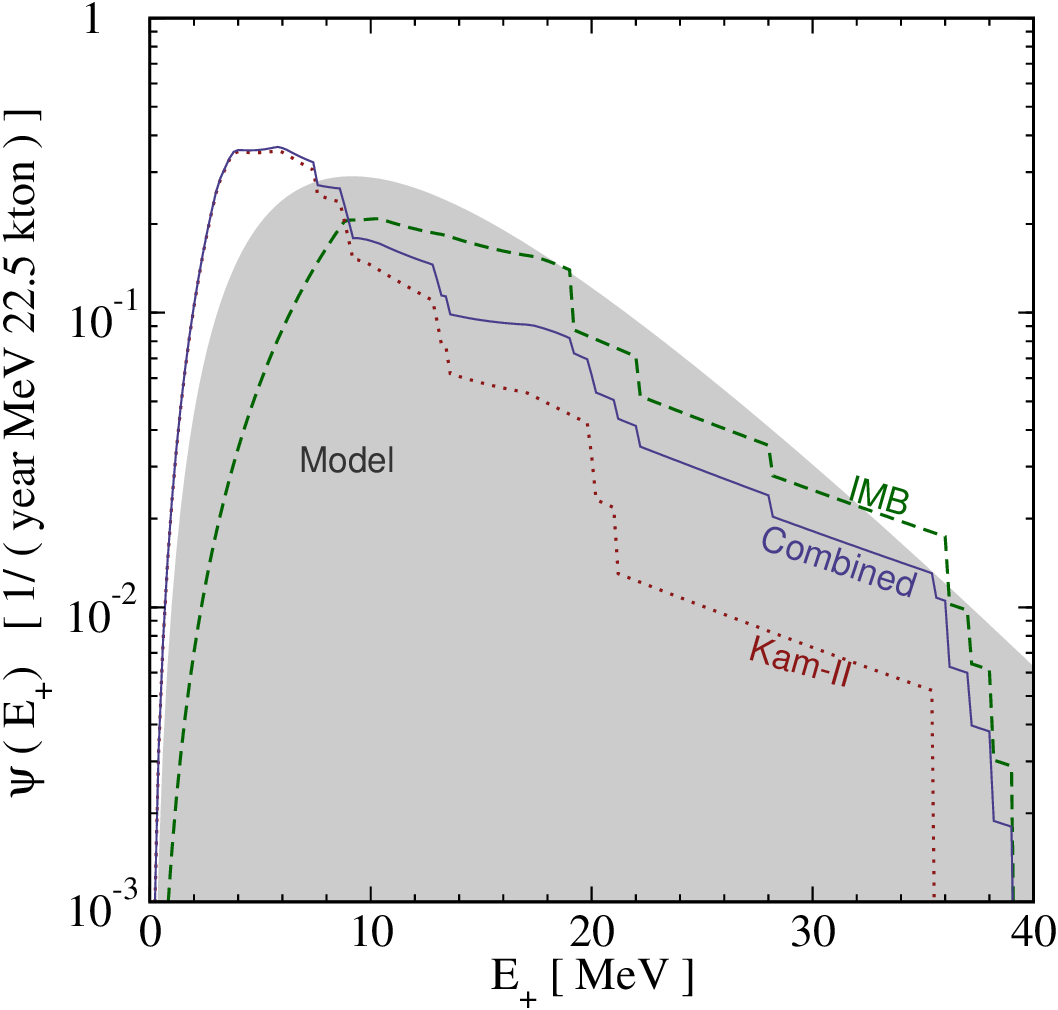}
\caption{\label{Fig:4}
The DSNB detection spectra based on the neutrino spectra inferred from either the Kam-II or IMB data sets alone or their combination, through the $\delta$-function prescription  which enables us to deduce the DSNB spectra without applying any smoothing (as discussed in the text), compared to a model (shaded shape) with canonical neutrino emission parameters.
}
\end{figure}

The neutrino emission per supernova times the supernova rate, when integrated over redshift and convolved with the cross section, yields the detected spectrum of DSNB neutrinos,
\begin{equation} 
\psi(E_+) =\frac{c}{H_0} \sigma(E_\nu) N_t \int_{0}^{z_{max}}  \phi(E_\nu[1+z])
\frac{{R}_{SN} (z)} {h(z)} dz \, ,
\label{eq:dsnb}
\end{equation}
where $c$ is the light speed and $H_0$ is the Hubble constant. For a modern detector like Super-Kamiokande (SK), the efficiency can safely be taken to be unity. We will present our results for a Super-Kamiokande sized detector of fiducial volume 22.5~kton, corresponding to $N_t = 1.5 \times 10^{33}$.

In Fig.~\ref{Fig:3}, the presented DSNB spectra are calculated by substituting
the smooth neutrino spectra inferred from either the Kam-II or IMB data sets alone or their combination as in Fig.~\ref{Fig:2}, into Eq.~(\ref{eq:dsnb}). A canonical neutrino emission spectrum described by Fermi-Dirac distribution ($E_0= 15$~MeV and $L= 5 \times 10^{52}$ ergs~\cite{Yuksel:2005ae}) is also presented. 

Note that we have replaced each detected positron  by a Gaussian to obtain a sufficiently smooth neutrino spectrum. This could inadvertently introduce some arbitrariness and distort the results.  Since the calculation of the DSNB already involves an integration over redshift, we can instead predict the DSNB directly from the data, which we call the $\delta$-function prescription. When redshifting is taken into account,  Eq.~(\ref{eq:phi}) can be cast as:
\begin{eqnarray} 
\phi(E_\nu[1+z])&=&
{4 \pi D^2} \sum_i \frac{\delta([1+z](E_++\Delta)-(E_i+\Delta))}{ N(E_{i}) \, \sigma(E_{i} +\Delta) }
\nonumber \\
&=&{4 \pi D^2} \sum_i \frac{\delta(z - (E_i-E_+) /E_\nu) }{E_\nu N(E_{i}) \, \sigma(E_{i} +\Delta) }
\end{eqnarray}
where we have used the delta function identity; $\delta(ax-b)=\delta(x-b/a)/a$. Substitution of this relation into Eq.~(\ref{eq:dsnb}) eliminates the integration over redshift, so that the detected positron spectrum is
\begin{equation} 
\psi(E_+) =\frac{c}{H_0} \sum_i 
\frac{\sigma(E_\nu) N_t \,{4 \pi D^2}  }{E_\nu N(E_{i}) \, \sigma(E_{i} +\Delta) }
\frac{{R}_{SN} ( (E_i-E_+) /E_\nu)} {h( (E_i-E_+) /E_\nu)}\, .
\label{eq:dsnbdelta}
\end{equation}

Figure~\ref{Fig:4} displays the resultant DSNB detection spectra, as deduced from the Kam-II or IMB data sets alone or their combination, and compares to the model. Apart from the very high energy tail, both the smooth method (as presented in Fig.~\ref{Fig:3}) and the $\delta$-function method agree quite well over all energy ranges. In Table~\ref{table:dsnb}, we summarize the DSNB event rates in various ranges of visible energy for both smooth and $\delta$-function prescriptions (all values are quoted per 22.5 kton per year). The values in the last column, reporting the event rates in the energy range of 18--26 MeV can be compared to the event rate limit of $\sim$2 at 95\% C.L. (as inferred in Ref.~\cite{Yuksel:2005ae} from the $\bar{\nu}_e$ flux limit of $\sim 1.2$ cm$^{-2}$ s$^{-1}$ at 95\% C.L. reported by SK~\cite{Malek:2002ns} through non-detection of excess counts above background fluctuations), suggesting that the DSNB is already tantalizingly close to detection. In the range 10--18 MeV, the event rates are nearly as large as the usual predictions based on supernova models, suggesting an imminent discovery of DSNB is well within the reach of current detectors, and especially promising if thresholds are reduced by the addition of gadolinium~\cite{Beacom:2003nk}.  

There are various advantages to predicting the DSNB directly from the data rather than  through the intermediate stages of fitting functions. The DSNB mainly consists of the higher energy part of the neutrino spectrum and is much less susceptible to uncertainties/backgrounds around the threshold of detectors. While theoretically motivated formulas attempt to explain the bulk of the detected events, they tend to put more weight in the low energy part at the expense of introducing distortions at the high energy parts of distributions relevant for the DSNB.  For a high threshold of 18 MeV, almost all of the statistical weight comes from the IMB data.  When the energy threshold is decreased, the Kam-II data will be given more weight. Note also that we do not attempt to deconvolve any poorly-understood physical effects from the observed spectrum to deduce the original spectra at formation. We are directly relating the observed spectrum of SN 1987A to the prediction for the observable DSNB spectrum from many supernovae.

\begin{table}[t]
\caption{The DSNB event rates in various ranges of visible energy from the
spectra displayed in Fig.~\ref{Fig:3} and Fig.~\ref{Fig:4}. All quoted values are per 22.5 kton per year.
\label{table:dsnb}}
\begin{ruledtabular}
\begin{tabular}{lcccccc}
Range (MeV) & \multicolumn{2}{c}{4-10} & \multicolumn{2}{c}{10-18} & \multicolumn{2}{c}{18-26}\\
& $\delta$& smooth& $\delta$& smooth& $\delta$ & smooth\\ 
\hline
Kam-II	& 1.8  & 1.7 & 0.7 & 1.0 & 0.2 & 0.3 \\
IMB     & 0.6  & 0.8 & 1.4 & 1.3 & 0.6 & 0.9 \\
\textbf{Combined}& \textbf{1.8} &\textbf{1.8} & \textbf{1.0} &\textbf{1.2} & \textbf{0.4} &\textbf{0.5}\\
Model	& --   & 1.5 & --  & 1.8 & -- & 0.8 
\end{tabular}
\end{ruledtabular}
\end{table}

The idea of normalizing the DSNB flux prediction to the SN~1987A data is appealingly empirical, and was first introduced by Fukugita and Kawasaki~\cite{Fukugita:2002qw}, and later developed in greater detail by Lunardini~\cite{Lunardini:2005jf}.  The disadvantage of relying on the SN~1987A data is the uncertainties and apparent disagreements of the sparse data.  While Fukugita and Kawasaki~\cite{Fukugita:2002qw} used only the IMB data, obtaining a neutrino emission per supernova in agreement with theoretical models, Lunardini showed that including the Kam-II data in the spectral fit leads to a more uncertain DSNB flux prediction, including the possibility of its being significantly lower than found by other authors.  Our predictions for the combined Kam-II and IMB data are consistent with those of Lunardini when differences in the star formation rate are taken into account.  However, as we show, it is the IMB data that are presently more relevant for the detectable DSNB flux, and accordingly, pessimistic DSNB detection rate predictions are disfavored.  The fact that 6 of the SN 1987A events were detected above 35 MeV strongly precludes a supernova spectrum that is either too soft or too low.

As with other forms of statistical inference, we are assuming that the observed data were representative of the truth (i.e., which is the maximally likely outcome), while considering how uncertain our subsequent results are.  We are not using theory or other considerations to judge that the data were subject to any particular statistical fluctuation. In particular, we found no evidence from the data alone that the Kam-II and IMB data sets were incompatible.  A second point to consider is whether SN 1987A was a representative supernova.  Despite the large range of progenitors, and the wide variety of optical supernova properties, the time-integrated neutrino signals from core-collapse supernovae are expected to be mostly uniform, as they are well-connected to the properties of the newly-produced neutron stars.  We note that both the SN 1987A data and the present SK limit on the DSNB flux both depend on statistical uncertainties at the $\sim 30\%$ level.  It may well be that SN 1987A is not a representative supernova, and indeed, this is part of what we want to test with the DSNB~\cite{Yuksel:2005ae}. As shown in Figs.~3 and 4, and summarized in Table~I, more typical results from theoretical expectations give a larger DSNB flux. Part of our point is that the SN 1987A data, when considered without theoretical priors, do not support a very low DSNB flux.

We now comment on the effects of considering also the data from the Baksan detector~\cite{Alekseev:1987ej,Alekseev:1988gp}. The relevant properties of this detector were very similar to that of Kam-II, except for being $\sim10$ times smaller. Thus using the Kam-II yield, $\sim1$ event would be expected in Baksan, while 5 were observed, and with a somewhat higher average energy than in Kam-II. If the Baksan data were a faithful representation of the true spectrum, then the corresponding DSNB flux would be enhanced by a factor $\gtrsim 5$, which is likely already excluded. The most likely resolution is that Baksan saw an upward fluctuation of their large background rate, perhaps along with some events from SN~1987A.

\section{Discussion and Conclusions}
Neutrinos play a crucial role in the life and death of massive stars and so far, they are the only messengers that enable us to probe the inner workings of a core collapse. Thus,  two decades after the detection of neutrinos from SN 1987A,  they continue to attract much attention. 

We assess the compatibility of the Kam-II and IMB data sets in a model-independent way and turn to the data directly, skipping any intermediate stages and assumptions about the incoming neutrino spectrum. Our main conclusion is that if one drops theoretical prejudices on the spectral shape, the Kam-II and IMB detections can be reconciled. The two data sets are primarily sensitive to different energy regimes of the neutrino spectrum (as seen in Fig.~\ref{Fig:1}), and only their proper combination probes the whole energy range. The actual tension is between the adopted theory and the observations, since assumed theoretical shapes are always limited to a certain subset of all mathematical possibilities which do not necessarily describe the true underlying physics.  When using the data directly, we consider the combined spectrum, based on both data sets, to be the most reliable.

Using parameter-free inferential statistical methods, we have shown that the combined Kam-II and IMB data can be best explained by a spectral shape that is enhanced both at the peak and the tail of the spectrum and depressed in between, compared to a well-known Fermi-Dirac spectrum (e.g., Fig.~\ref{Fig:2}). Such a distribution may be arising from many different physical processes, including the details of core collapse or neutrino mixing, which are beyond the scope of this study. Once the effective received neutrino spectrum is determined, one can then work backwards and uncover the scenarios that will yield this measured spectrum. Considering that supernova models still fail to robustly explode despite increased sophistication in modeling, it is an alluring possibility that a key element may still be missing~\cite{Buras:2003sn}. The model theoretical spectra with canonical emission parameters adopted in many studies could provide and capture essentials of the supernova neutrino spectrum, and be adequate for most purposes. While theoretical models still provide essential guidance, the necessity of fresh data on supernova neutrinos is obvious.  The rarity of galactic supernovae, with the most optimistic rate estimates of at most a few per century, makes this a challenge.

Apart from the proposal to detect individual  neutrinos  from galaxies within the 10 Mpc neighborhood of the Milky Way with future Mton-scale detectors~\cite{Ando:2005ka}, the DSNB presents the only sensible alternative to gain information on neutrino emission from  supernovae. We use a nonparametric approach in order to determine one observable, the Diffuse Supernova Neutrino Background, directly from another observable, the SN~1987A data on supernova neutrino emission,  rather then proceeding through intermediate stages of fitting functions. We show that this prediction cannot be too small (especially in the 10-18 MeV range), since the majority of the detected events from SN~1987A were above 18 MeV (with 6 above 35~MeV). We emphasize that our DSNB predictions are not very dependent on the details of our analysis procedure, as can be seen by comparing Figs.~\ref{Fig:3} and \ref{Fig:4}, and especially by comparing the ``$\delta$" and ``smooth" cases in Table~I.

These results are nearly as large as the usual predictions based on supernova models, suggesting an imminent discovery of DSNB is well within the reach of current detectors, and especially promising if thresholds are reduced by the addition of gadolinium~\cite{Beacom:2003nk}. A gadolinium-enhanced Super-Kamiokande should also be able to provide a measurement of the spectral shape~\cite{Yuksel:2005ae,Lunardini:2006pd}, providing further clues. The DSNB, which may even one day serve as a tool for extracting cosmological evolution parameters~\cite{Hall:2006br},  is a leading candidate along with the other contenders, like cosmogenic neutrinos~\cite{Yuksel:2006qb}, to open new doors to the cosmos and provide the first neutrino detection originating from cosmological distances.

\section*{ACKNOWLEDGMENTS}
We thank Arnon Dar and Georg Raffelt for helpful discussions, and Matt Kistler and Cecilia Lunardini for the same and also a careful reading of the manuscript.  This work is supported by the National Science Foundation under CAREER Grant No. PHY-0547102 to JB, and CCAPP at The Ohio State University.

\end{document}